\documentclass{webofc}
\usepackage[varg]{txfonts}   % Web of Conferences font
\usepackage[dvipsnames]{xcolor}
\usepackage[colorlinks=true,linkcolor=black,urlcolor=blue,citecolor=blue]{hyperref}

\begin{document}
\title{Far-off-equilibrium early-stage dynamics in high-energy nuclear collisions}
%
% subtitle is optionnal
%
%%%\subtitle{Do you have a subtitle?\\ If so, write it here}

\author{\firstname{Chandrodoy} \lastname{Chattopadhyay}\inst{1}\fnsep\thanks{\email{cchatto@ncsu.edu}} \and
        \firstname{Ulrich} \lastname{Heinz}\inst{2}\fnsep\thanks{\email{heinz.9@osu.edu}} \and
        \firstname{Thomas} \lastname{Sch\"afer}\inst{3}\fnsep\thanks{\email{tmschaef@ncsu.edu}}
        % etc.
}

\institute{Department of Physics, North Carolina State University, Raleigh, NC 27695, USA
\and
           Department of Physics, The Ohio State University, Columbus, OH 43210, USA
\and
           Department of Physics, North Carolina State University, Raleigh, NC 27695, USA
          }

\abstract{%
  We explore the far-off-equilibrium aspects of the (1+1)-dimensional early-stage evolution of a weakly-coupled quark-gluon plasma using kinetic theory and hydrodynamics. For a large set of far-off-equilibrium initial conditions the system exhibits a peculiar phenomenon where its total equilibrium entropy decreases with time. Using a non-equilibrium definition of entropy based on Boltzmann's H-function, we demonstrate how this apparently anomalous behavior is consistent with the second law of thermodynamics. We also use the H-function to formulate `maximum-entropy' hydrodynamics, a far-off-equilibrium macroscopic theory that can describe both free-streaming and near-equilibrium regimes of quark-gluon plasma in a single framework.  
  
}
\maketitle
\vspace*{-5mm}
\section{Introduction}
\label{sec1}
\vspace*{-2mm}

Precise determination of transport coefficients like the specific viscosities $\eta/s$ and $\zeta/s$ of the quark-gluon plasma formed in high-energy nucleus-nucleus collisions hinges upon accurately modeling the stress tensor ($T^{\mu\nu}$) evolution during the system's early stage. This stage is characterised by far-off-equilibrium dynamics which may be modeled by weakly coupled kinetic theory until ${\cal O}(1)$\,fm/$c$ \cite{Kurkela:2018wud, Liyanage:2022nua}. This approach is, however, numerically daunting as solving kinetic theory amounts to tackling a 7-dimensional problem in phase-space. Moreover, if one is only interested in the evolution of macroscopic quantities like $T^{\mu\nu}$, solving for the full kinetic distribution is likely unnecessary. It is thus desirable to have a macroscopic framework which can model the far-off-equilibrium evolution of $T^{\mu\nu}$ both physically accurately and numerically efficiently. In this work, we first explore the sensitivity of the $T^{\mu\nu}$ evolution in kinetic theory to initial state momentum anisotropies of the plasma. By considering extreme off-equilibrium initial conditions for a quark-gluon gas undergoing Bjorken expansion \cite{Bjorken:1982qr}, we point out non-intuitive out-of-equilibrium effects arising in kinetic theory. In the second part we formulate a new macroscopic theory (ME-hydrodynamics) which can be used to describe in a single framework both the far-off-equilibrium pre-hydrodynamic and the near-equilibrium dissipative hydrodynamic regimes of the plasma.  

\vspace*{-3mm}
\section{Kinetic theory of a massive quark-gluon gas}
\label{sec2}
\vspace*{-2mm}

\begin{figure}[b]
\centering
\sidecaption
\includegraphics[width=9cm,clip]{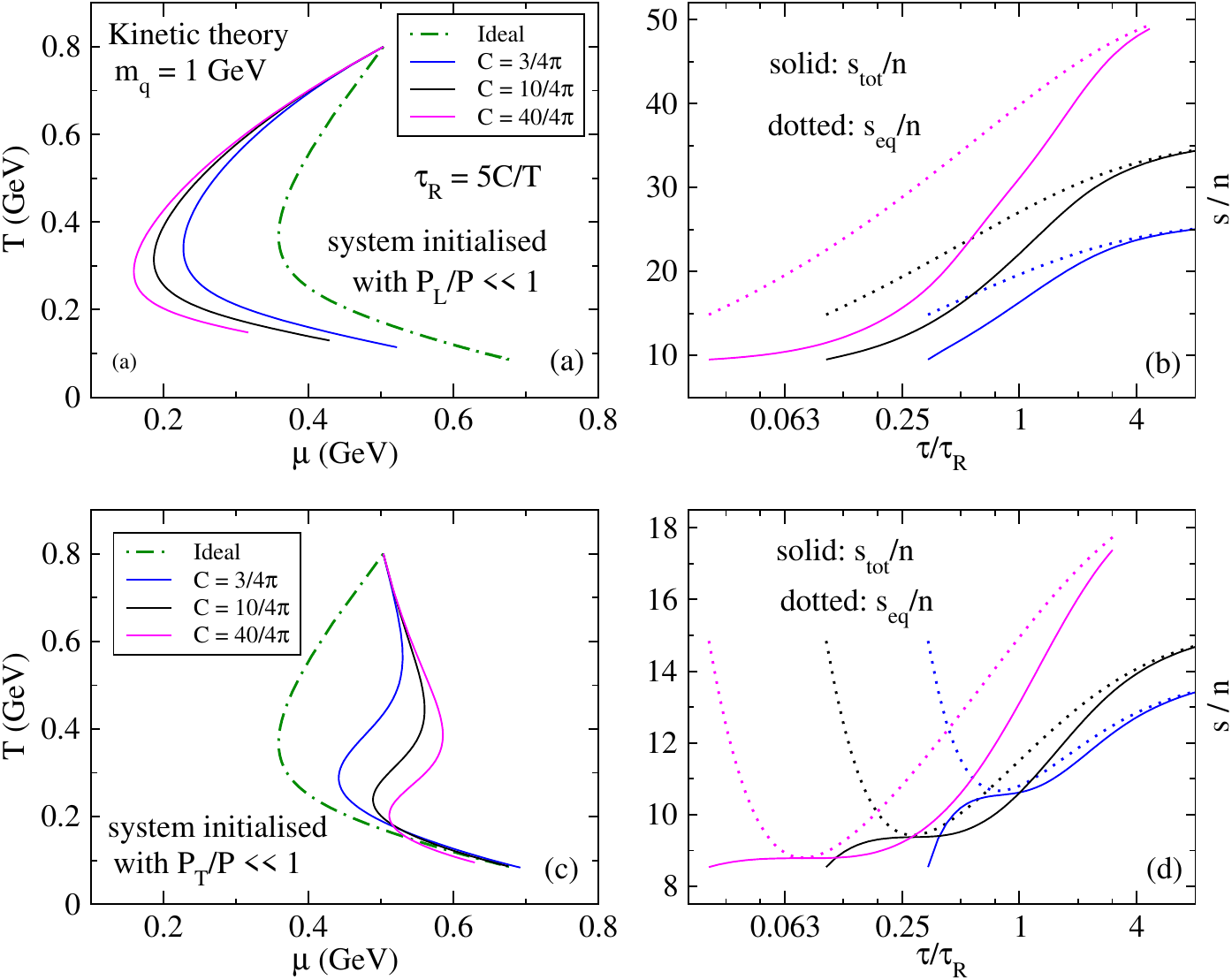} 
\caption{
Phase trajectories and corresponding entropy evolution of a quark-gluon gas initialized near the {\it stable} fixed point of early-time Bjorken dynamics, i.e., $P_L/P{\,\ll\,}1$ \cite{Blaizot:2021cdv} (upper panels), and the {\it unstable} fixed point $P_T/P{\,\ll\,}1$ (lower panels). The magenta curves depict a very weakly interacting gas whereas the green curve $(\tau_R \to 0)$ is for a strongly interacting system (perfect fluid).
}
\label{fig1}
\end{figure}

\begin{figure}
\centering
\sidecaption
\includegraphics[width=4.4cm,clip]{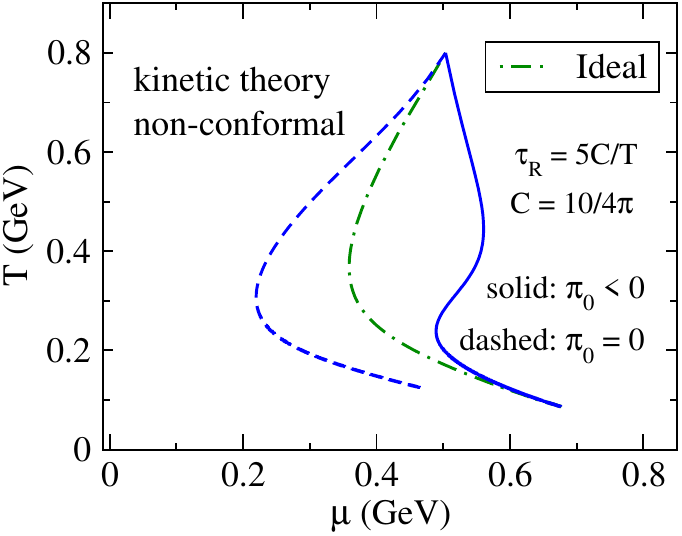} \,\,\,
\includegraphics[width=4.5cm,clip]{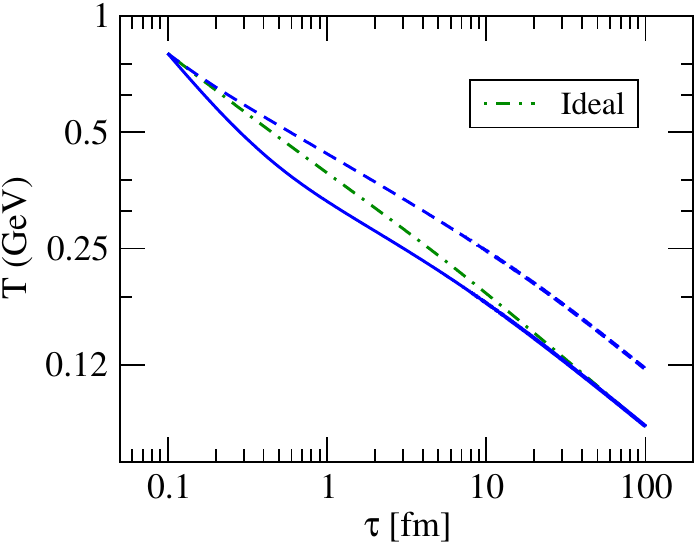} 
\caption{Non-equilibrium cooling: the solid (dashed) blue curve, initialised with non-equilibrium (equilibrium) initial conditions, cools more rapidly (slowly) than a dissipationless system. 
}
\label{fig2}
\end{figure}

For a weakly interacting gas of quarks, anti-quarks, and gluons undergoing boost-invariant Bjorken expansion along the beam axis, we solve the Boltzmann equation in a relaxation-time approximation,
\begin{equation}
    \frac{\partial f^i}{\partial \tau} = - \frac{1}{\tau_R(T)} \, \left(f^i - f^i_{\mathrm{eq}} \right). \label{RTA_BE}
\end{equation}
 Here $\tau_R$ is the microscopic relaxation time, and the superscript $i{\,\in\,}\{q, \bar{q}, g\}$ on the kinetic distributions distinguishes between particle species. $f^i_{\mathrm{eq}}$ are given by Fermi-Dirac (for quarks and anti-quarks) or Bose-Einstein (for gluons) distributions which involve the Landau matched effective temperature and quark chemical potential $(T, \mu)$. Symmetries of Bjorken flow imply vanishing net-quark diffusion, i.e. $n(\tau){\,\propto\,}1/\tau$ and $T^{\mu\nu} = \sum_i \int_{p_i} p_i^\mu p_i^\nu f^i = \mathrm{diag}(e, P_T, P_T, P_L)$, where $P_T$ and $P_L$ are effective transverse and longitudinal pressures. A physical quantity that is of particular interest is the non-equilibrium entropy density (in the fluid rest frame), obtained from Boltzmann's H-function: 
\begin{align}
\label{entropy}
    s = - \sum_{i} \int_{p_i} \, \left( u \cdot p_i \right)\, \left[ f^i \ln f^i - \frac{1 + a_i \, f^i}{a_i} \ln\left( 1 + a_i \, f^i \right) \right], 
\end{align}
where $a_{q,\bar{q}} = -1$ and $a_g = 1$. In equilibrium $s\to s_{\mathrm{eq}} = (e + P - \mu n)/T$. In Fig.~\ref{fig1} we show solutions of kinetic theory for two sets of extreme far-off-equilibrium initial conditions (see figure caption) which were set up using a Romatschke-Strickland (RS) distribution \cite{Romatschke:2003ms, Chattopadhyay:2021ive}. Although all curves start with the same effective $(T, \mu_B)$, the phase trajectories are quite sensitive to the choice of initial momentum space anisotropy. In Bjorken flow, Navier-Stokes hydrodynamics predicts that the ratio $s_\mathrm{eq}/n$ must increase over time due to viscous heating. While this is indeed the case for panel (a) (see dotted lines for $s_\mathrm{eq}/n$ evolution in (b)), this expectation is not borne out for the trajectories in panel (c). Here, $s_{eq}/n$ decreases for a certain duration of time. However, this does not imply a violation of the second-law of thermodynamics as the total entropy per baryon which includes non-equilibrium effects never decreases. The feature of decreasing equilibrium entropy per baryon density results in a peculiar phenomena which we call `non-equilibrium cooling' (see Fig. \ref{fig2}). Here, the effective temperature falls even faster than what is expected for an ideal (inviscid) fluid.     

\vspace*{-2mm}
\section{Maximum-entropy truncation of the Boltzmann equation}
\label{sec3}
\vspace*{-1mm}

The Boltzmann equation can be expressed as an infinite hierarchy of equations for momentum moments of $f(x,p)$ \cite{Denicol:2012cn} where low-order moments corresponding to components of $T^{\mu\nu}$ are coupled to higher-order `non-hydrodynamic' moments. To obtain a macroscopic description solely in terms of $T^{\mu\nu}$, the infinite hierarchy has to be truncated by expressing the non-hydrodynamic moments in terms of an approximate kinetic distribution using only information contained in $T^{\mu\nu}$. Based on Jaynes's insights on the connections between statistical mechanics and information theory \cite{Jaynes:1957zza}, Everett {\it et al.} \cite{Everett:2021ulz} recently proposed a novel way of reconstructing a kinetic distribution from the energy-momentum tensor using the maximum entropy principle. The idea is to find the distribution function $f(x,p)$ that maximizes the non-equilibrium entropy density (\ref{entropy}), 
subject to the information (constraint) that $f(x,p)$ reproduces the given 10 components of $T^{\mu\nu}$. For a single component gas the maximum entropy distribution is \cite{Everett:2021ulz}
\begin{align}
    f_{\mathrm{ME}}(x,p) = \left[\exp\left( \frac{\Lambda_{\mu\nu} p^\mu p^\nu}{u \cdot p} \right) - a \right]^{-1}, 
\end{align}
where $\Lambda_{\mu\nu}$ are Lagrange multipliers corresponding to $T^{\mu\nu}$. Landau matching conditions further simplify the argument of the exponential \cite{Chattopadhyay:2022sxk}. Unlike the commonly used distributions for Grad or Chapman-Enskog (CE) truncation, $f_{\mathrm{ME}}$ is positive definite for all momenta and allows for non-equilibrium matching to conserved currents for a wide range of non-equilibrium stresses. It also  ensures that the resulting macroscopic framework, which we call ME-hydro, has a non-negative entropy production rate \cite{Calzetta:2010au}, and that in the limit of small viscous stresses ME-hydro reduces to second-order Chapman-Enskog fluid dynamics \cite{Everett:2021ulz}.

\vspace*{-3mm}
\section{ME-hydro vs. RTA kinetic theory in Bjorken and Gubser flows}
\label{sec4}
\vspace*{-1mm}

The exact evolution equations for the 3 independent components of $T^{\mu\nu}{\,=\,}\mathrm{diag}(e, P_T, P_T, P_L)$ in Bjorken flow are given by
\begin{align}
\frac{de}{d\tau} = - \frac{e + P_L}{\tau}, \,\, \, \, 
\frac{dP_T}{d\tau} = - \frac{P_T - P}{\tau_R} - \frac{P_T}{\tau} + \frac{\zeta_T}{\tau}, \,\,\,
\frac{dP_L}{d\tau} = - \frac{P_L - P}{\tau_R}  - \frac{3 P_L}{\tau} +  \frac{\zeta_L}{\tau}\,. 
\end{align}
The terms $(\zeta_T, \zeta_L)$ introduce couplings to `non-hydrodynamic' moments of $f(\tau, p_T, p_z)$; for example, $\zeta_L = \int_p \, E_p^{-2} \, p_z^4 \, f$. To truncate we replace $f \mapsto f_{\mathrm{ME}}$ where $f_{\mathrm{ME}}$ is constructed using the instantaneous values of $(e, P_T, P_L)$ \cite{Chattopadhyay:2023hpd}. % 
\begin{figure}[hbt!]
\centering
\sidecaption
\includegraphics[scale=0.31]{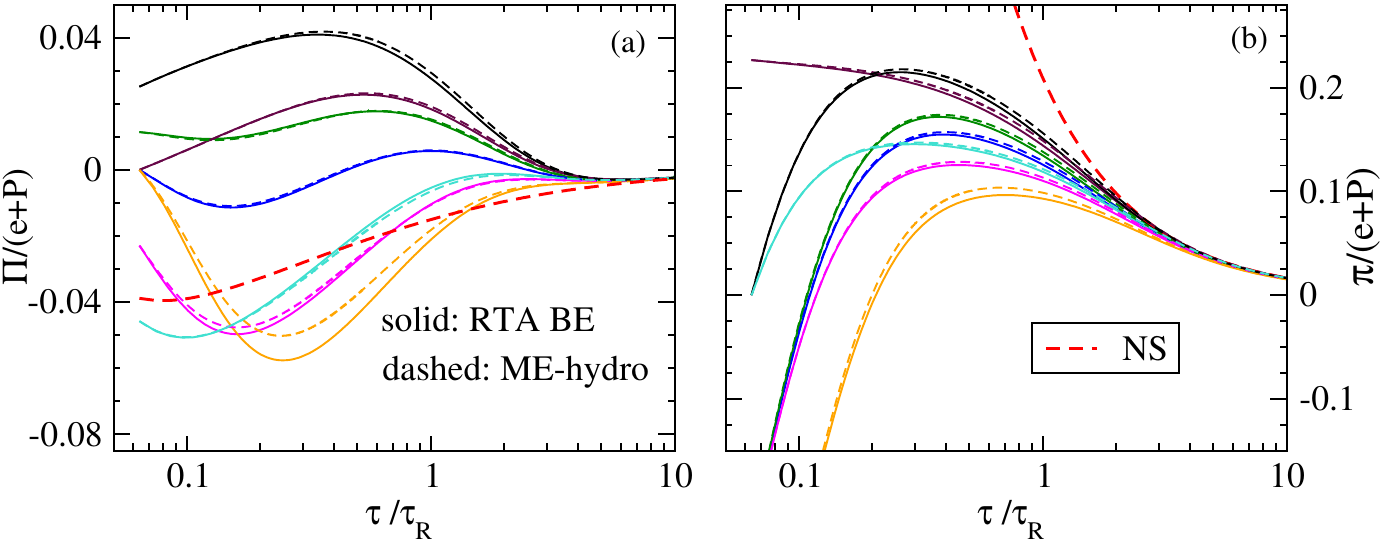}\!\!\! 
\caption{Comparison of ME-hydro results with non-conformal RTA BE for the scaled bulk viscous pressure $\Pi{\,=\,}(P_L{+}2 P_T{-}3P)/3$ (left panel) and shear stress tensor compo- nent $\pi{\,=\,}2 (P_T{-}P_L)/3$ (right).  
The red- dashed curves are first-order hydrodynamic predictions (Navier-Stokes). 
}
\vspace*{-2mm}
\label{fig3}
\end{figure}%
In Gubser flow \cite{Gubser:2010ui} the exact evolution equations for the two independent (dimensionless) variables $(\hat{e}, \hat{P}_T)$ as functions of de-Sitter `time' $\rho$
are similarly truncated using $f_{\mathrm{ME}}$ \cite{Chattopadhyay:2023hpd}. Figures~\ref{fig3}-\ref{fig4} show that ME-hydro is in excellent agreement 
\begin{figure}[htb!]
\centering
\sidecaption
\includegraphics[width=4.5cm,clip]{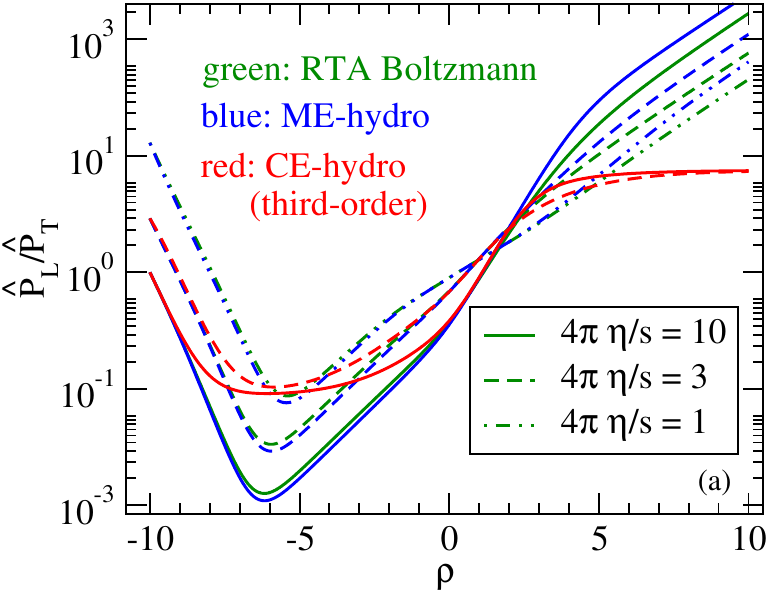} \,\,\,
\includegraphics[width=4.5cm,clip]{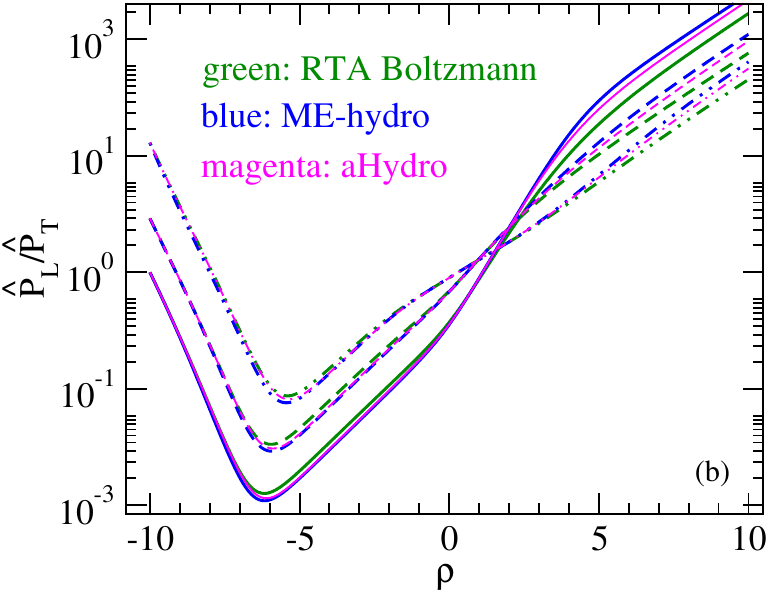} 
\caption{Evolution of the pressure anisotropy in Gubser flow. At early and late times the system approaches longitudinal and transverse free- streaming regimes, respectively.}
\label{fig4}
\end{figure}%
with the underlying kinetic theory for both of these profiles even when the system is far-off-equilibrium. Figure~\ref{fig4}a shows that Chapman-Enskog hydrodynamics \cite{Chattopadhyay:2018apf} fail to capture the late-time transverse free-streaming regime of Gubser flow. The only framework that performs slightly better than ME-hydro is anisotropic hydrodynamics \cite{Martinez:2010sc, Martinez:2017ibh} (shown in panel (b)) which uses the RS ansatz as a truncation distribution. Further (numerical) analysis will determine which of these two frameworks better captures the far-off-equilibrium aspects of kinetic theory in generic flows without the restrictive symmetry constraints of Bjorken and Gubser flows.\\[-1.5ex]

\noindent{\bf Acknowledgements:} Discussions with J. P. Blaizot and S. Jaiswal are gratefully acknowledged. This work was supported by the US Department of Energy, Office of Nuclear Physics under contracts 
\rm{DE-FG02-03ER41260} (CC, TS) and \rm{DE-SC0004286} (UH), as well as by the Ohio State University Emeritus Academy (UH).
\vspace*{-3mm}

%
% BibTeX or Biber users please use (the style is already called in the class, ensure that the "woc.bst" style is in your local directory)
 \bibliography{references.bib}

\end{document}